\documentclass[11pt,twoside]{article}
\usepackage{graphicx,epsfig,natbib,epstopdf}
\usepackage{CS18}
\begin{document}
%
%
%
\title{The coolest 'stars' are free-floating planets}
%
%
\author{V. Joergens$^{1,2}$, M. Bonnefoy$^{3}$, Y. Liu$^{4}$
A. Bayo$^{1}$, S. Wolf$^{5}$}

\affil{$^1$Max-Planck Institut f\"ur Astronomie, K\"onigstuhl~17, 69117 Heidelberg, Germany}
\affil{$^2$ 	Institut f\"ur Theoretische Astrophysik,
        Zentrum f\"ur Astronomie der Universit\"at Heidelberg,
	Albert-Ueberle-Str. 2, 69120 Heidelberg, Germany}
\affil{$^3$ Institut de Plan\'etologie et d'Astrophysique de Grenoble, UMR 5274, 
Grenoble, 38041, France}
\affil{$^4$ Purple Mountain Observatory \& Key Laboratory for Radio Astronomy, 
        Chinese Academy of Sciences, Nanjing 210008, China}
\affil{$^5$ Institut f\"ur Theoretische Physik und Astrophysik,
	Universit\"at Kiel,  
        Leibnizstr. 15, 24118 Kiel, Germany}

\begin{abstract}
We show that the coolest known object that is probably formed in a star-like mode is 
a free-floating planet. We discovered recently that the free-floating 
planetary mass object OTS\,44 (M9.5, $\sim$12 Jupiter masses, age $\sim$2\,Myr) has 
significant accretion and a substantial disk. This demonstrates that the 
processes that characterize the canonical star-like mode of formation apply 
to isolated objects down to a few Jupiter masses.
We detected in VLT/SINFONI spectra that OTS\,44 has strong, broad, and variable 
Paschen $\beta$ emission. This is the first evidence for active accretion 
of a free-floating planet. The object allows us to study accretion and 
disk physics at the extreme and can be seen as free-floating analog of 
accreting planets that orbit stars. Our analysis of OTS\,44 shows that the 
mass-accretion rate decreases continuously from stars of several solar 
masses down to free-floating planets. 
We determined, furthermore, the disk mass (10 Earth masses) and further disk properties of OTS\,44 through 
modeling its SED including Herschel far-IR data.
We find that objects between 14 and 0.01 solar masses have the same ratio of the 
disk-to-central-mass of about 1\%. 
Our results suggest that OTS\,44 is formed like a star and that
the increasing number of young free-floating planets and ultra-cool  
T and Y field dwarfs are the low-mass extension of the stellar population. 
\end{abstract}

\section{Introduction}

We have witnessed in the last decade the detection of free-floating substellar objects
with increasingly lower masses and cooler temperatures that reach the canonical
planetary regime of a few Jupiter masses and a few hundred Kelvin. 
Examples are free-floating 
planetary mass objects in star-forming regions 
(e.g., \citealt{2000Sci...290..103Z}) and ultracool Y dwarfs in the field
(e.g., \citealt{2011ApJ...743...50C}).
One of the main open questions in the theory of star formation is: 
How do these free-floating planetary-like objects - and brown dwarfs in general - form? 
A high-density phase is necessary for the gravitational 
fragmentation to create very small Jeans-unstable cores. 
Proposed scenarios to prevent a substellar core in a dense environment
from accreting to stellar mass are ejection of the core through dynamical interactions
(e.g., \citealt{2001AJ....122..432R})
or photo-evaporation of the gas envelope through radiation of a nearby hot star
(e.g., \citealt{2004A&A...427..299W}).
Also suggested was the formation of brown dwarfs by disk instabilities in the outer regions 
of massive circumstellar disks (e.g., \citealt{2009MNRAS.392..413S}). 
Alternatively, brown dwarfs could form in a more isolated mode by direct collapse. 
For example, filament collapse (e.g., Inutsuka \& Miyama 1992) 
might form low-mass cores that experience high self-erosion
in outflows and become brown dwarfs (Machida et al. 2009).
A key to understanding star and brown dwarf formation is to observationally define 
the minimum mass that the canonical star-like mode of formation
can produce by detecting and 
exploring the main features characteristic of this process,
such as disks, accretion, outflows, and orbiting planets, for the lowest-mass objects.

The first indications for the existence of free-floating 
planetary mass objects came from observations in the very young 
$\sigma$ Orionis cluster
\citep{2000Sci...290..103Z}.
While it is difficult to establish the membership to $\sigma$ Ori and, therefore, the age and mass,
some very promising candidates were detected. An example is S\,Ori\,60, for which 
mid-IR excess indicates a young age \citep{2008ApJ...688..362L}.
Other candidates, however, turned out not to be members of the cluster,
such as S\,Ori\,70
(\citealt{2004ApJ...604..827B}; Pe\~na Ramirez, this volume).
The coolest known free-floating objects are Y dwarfs, of which 
17 are spectroscopically confirmed to date \citep{2014AJ....147..113C}. 
While we can expect many planetary mass objects among Y dwarfs,
the age and mass of these nearby field objects are in most cases unknown.
The first confirmed Y dwarf, WISE\,1828+2650 \citep{2011ApJ...743...50C},
was suggested to have a temperature and mass of only about 300\,K and 3-6\,$M_{\rm Jup}$, 
respectively, although there are indications that the temperature might be significantly larger 
\citep{2013Sci...341.1492D}.
Very recently a remarkably cold and low-mass object was detected in 
the very near vicinity of our Sun:  
WISE\,0855-0714 at a distance of 2.2\,pc has a proposed temperature and mass of
250\,K and 3-10\,$M_{\rm Jup}$, respectively \citep{2014ApJ...786L..18L}.
Nearby objects that do allow constraints on their age and, therefore, mass are members of 
young moving groups (see also Faherty, this volume). 
One of them, PSO\,J318.5-22, was recently detected to be a planetary-mass 
member of $\beta$ Pic. Depending on the assumed age for $\beta$ Pic
its mass is estimated to
6.5\,$M_{\rm Jup}$ (age: 12\,Myr) and 8.4\,$M_{\rm Jup}$ (21\,Myr), respectively
(\citealt{2013ApJ...777L..20L}; Allers, this volume).

To address the question how free-floating planets form, we observed OTS\,44, the 
lowest mass known member of the Chamaeleon~I star-forming region (Cha\,I, $\sim$2\,Myr, 160\,pc). 
OTS\,44 was first identified as a brown dwarf candidate in a deep near-IR imaging survey
\citep{1999ApJ...526..336O} and later
confirmed as very low-mass substellar object of spectral-type M9.5 
based on low-resolution near-IR and optical spectra 
\citep[e.g.][]{2004ApJ...617..565L}.
Its mass was estimated using near-IR spectra to lie in or very close to
the planetary regime ($\sim$6-17\,M$_{\rm{Jup}}$, \citealt{2013A&A...555A.107B}) 
with an average rounded value of 12\,M$_{\rm{Jup}}$.

\section{OTS\,44 - a 12 Jupiter mass object with a substantial disk}

Mid-IR photometry of OTS\,44 by Spitzer/IRAC and MIPS \citep{2008ApJ...675.1375L} 
indicated the presence of circumstellar material surrounding OTS\,44.
Recently, the disk of OTS\,44 was detected 
at far-IR wavelength by Herschel/PACS
\citep{2012ApJ...755...67H}. We re-analyzed the Herschel flux measurement \citep{joergens2014}
and find a slightly smaller value for OTS\,44 than \citet{2012ApJ...755...67H}.

\begin{figure}[h]
\centering
\includegraphics[width=85mm]{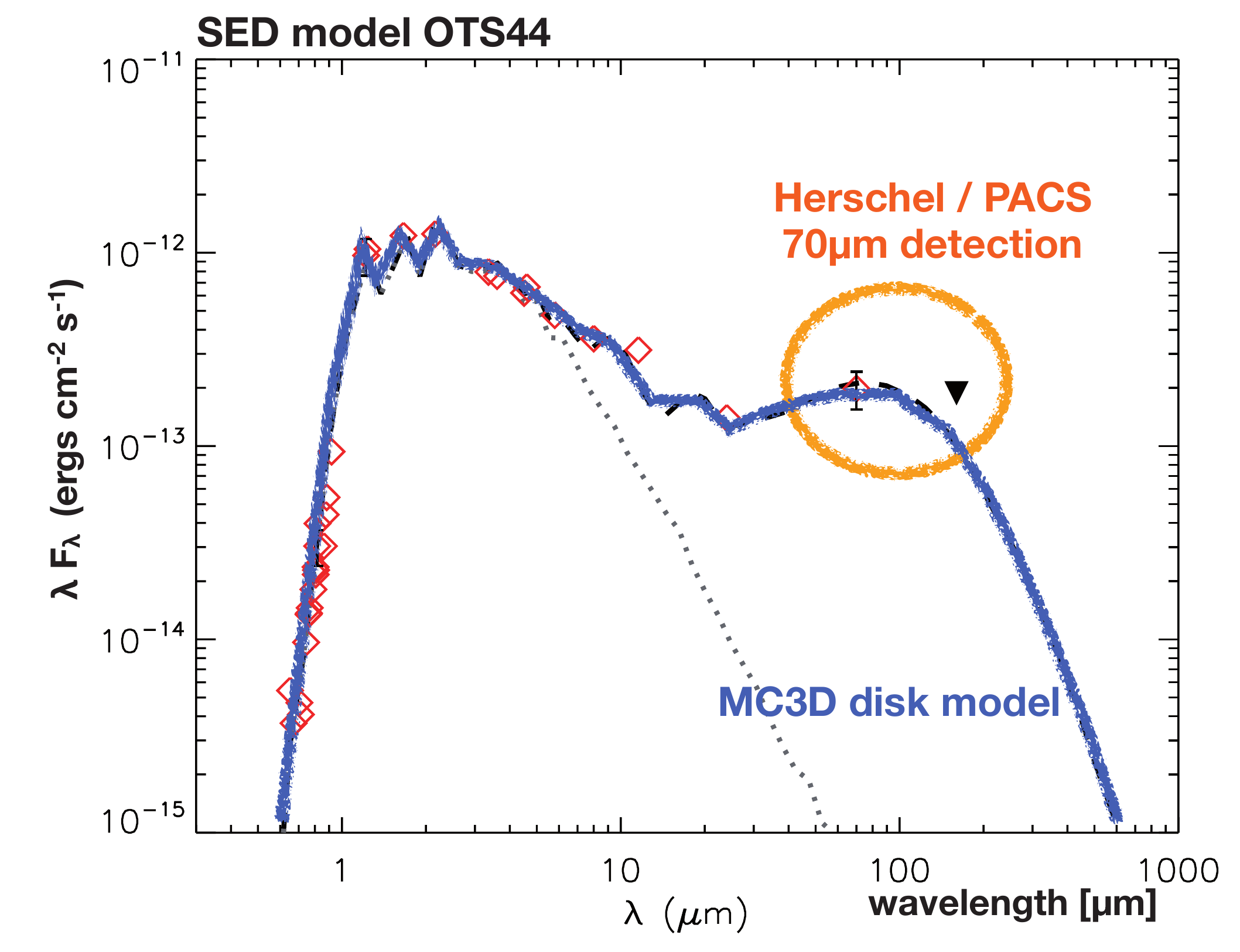}
\caption{
\label{fig:sed}
SED of OTS\,44. Shown are photometric measurements (red diamonds)
an upper limit for the 160\,$\mu{\rm{m}}$ flux (black triangle),
            the best-fit SED model (blue thick line), and 
the input BT-Settl photosphere model (gray dotted line).
This is a slightly revised version of the SED model of OTS\,44 in
\citet{2013A&A...558L...7J} using a revised Herschel flux measurement
\citep{joergens2014}.}
\end{figure}

We modeled the SED of OTS\,44 based on flux measurements from the optical to the far-IR
using the radiative transfer code \texttt{MC3D} \citep{2003ApJ...582..859W}.
We employ a passive-disk model consisting of a central substellar source 
surrounded by a parameterized flared disk in which dust and gas are well mixed and homogeneous throughout the system. 
We assume dust grains of astronomical silicate (62.5\%) and 
graphite (37.5\%) with minimum and maximum grain sizes of 0.005\,$\mu{\rm m}$ and 0.25\,$\mu{\rm m}$, respectively.
As there are model degeneracies between different 
disk parameters, we conducted a Bayesian analysis to estimate the validity range 
for each parameter. 
See \citet{2013A&A...558L...7J} for details.
Fig.\,\ref{fig:sed} shows the SED and the best-fit disk model of OTS\,44 based on the 
re-analysis of the Herschel fluxes \citep{joergens2014}.
We find that OTS\,44 has a highly flared disk ($\beta>$1.3)
with a disk mass of 3.25$\times10^{-5}\,M_{\odot}$, i.e. about $10\,M_{Earth}$.

%

\section{OTS\,44 - a 12 Jupiter mass object with significant accretion}

\begin{figure}
\centering
\includegraphics[height=.253\linewidth,clip]{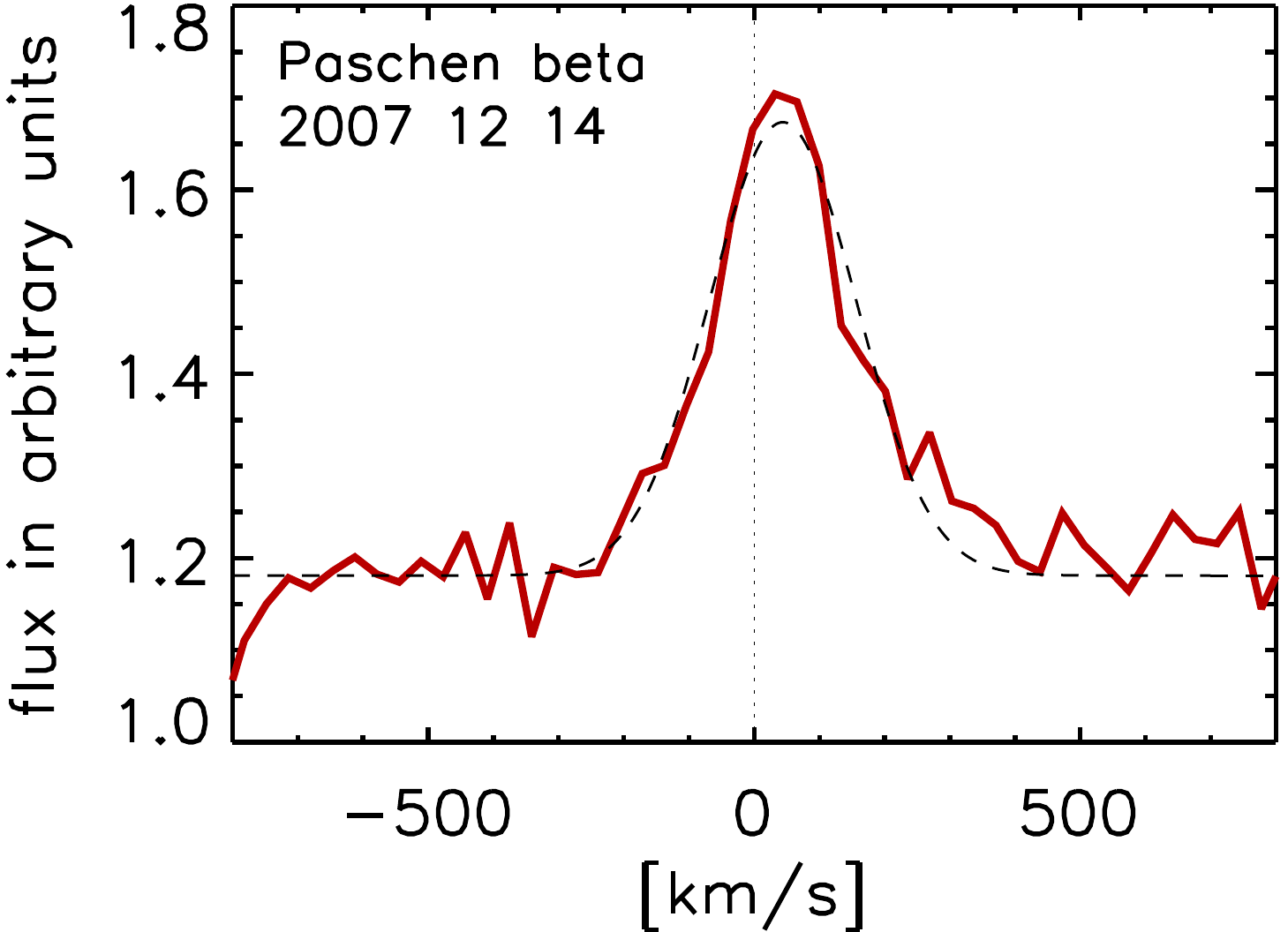}
\includegraphics[height=.253\linewidth,clip]{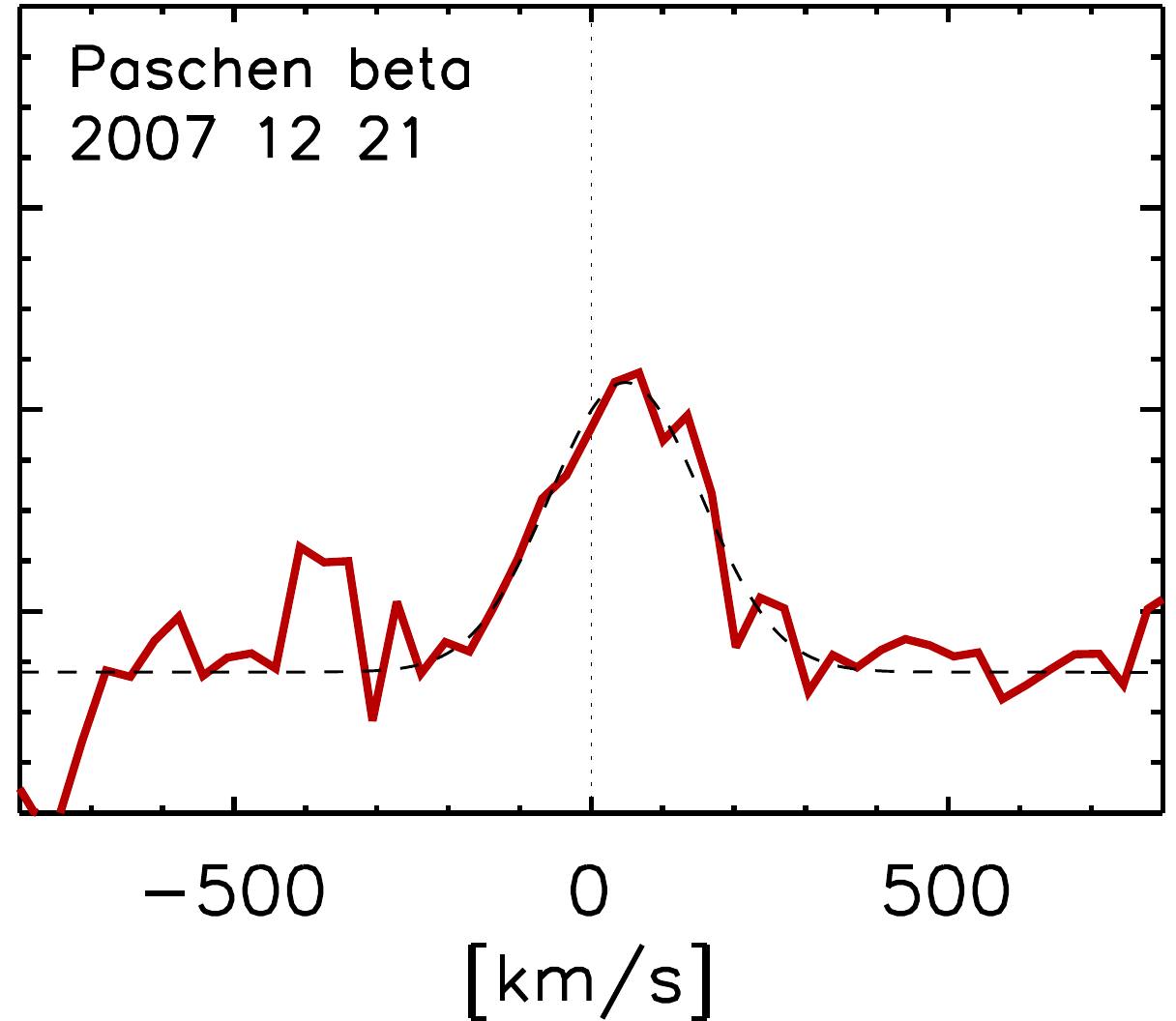}
\includegraphics[height=.253\linewidth,clip]{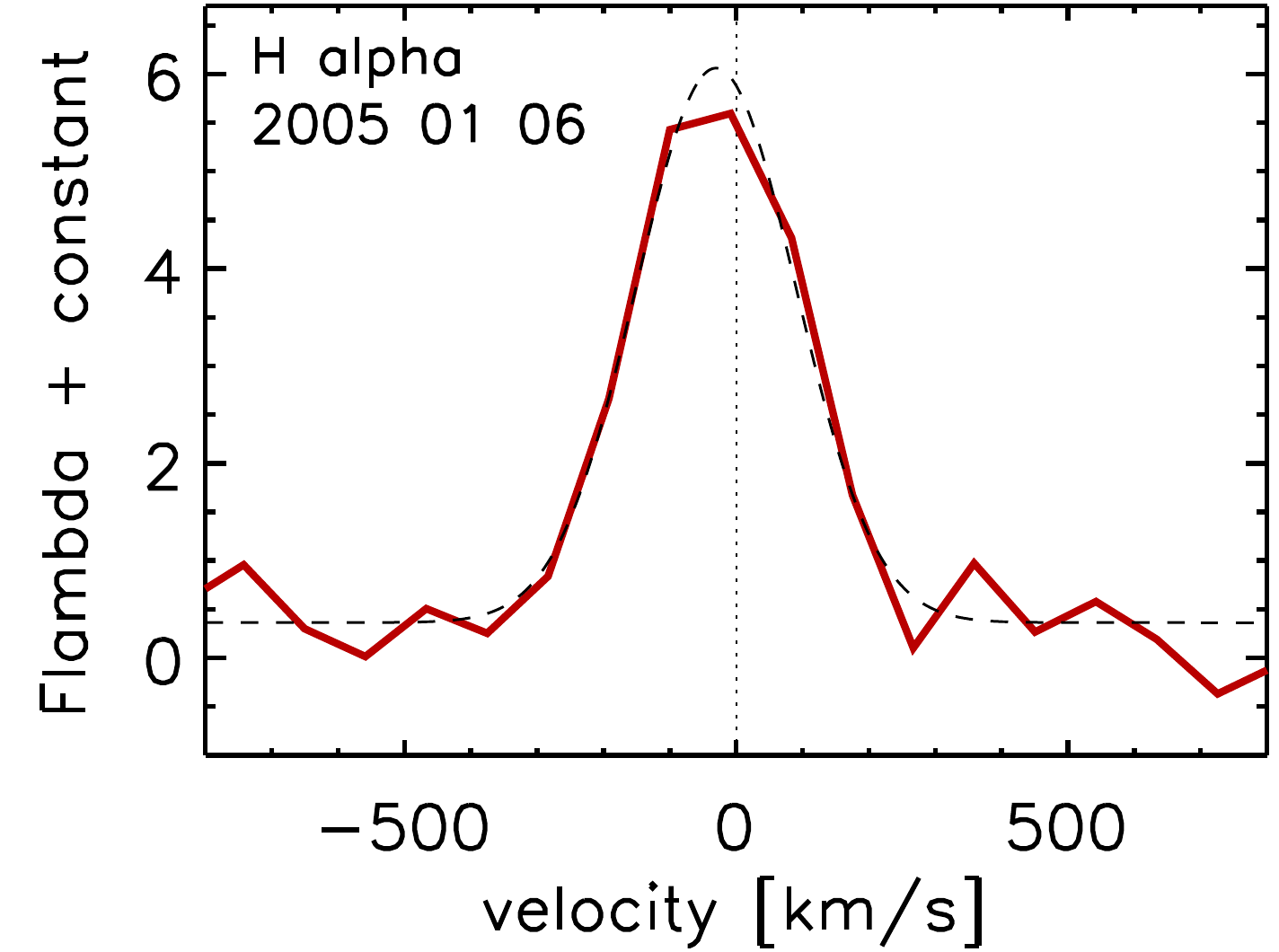}
\caption{
\label{fig:pabeta}
Pa\,$\beta$ emission of OTS\,44 in VLT/SINFONI spectra (left, middle)
and H$\alpha$ emission of OTS\,44 in a MAGELLAN\,I/IMACS spectrum 
(right).
The dashed lines are Gaussian fits to the profiles. 
From \citet{2013A&A...558L...7J}.
}
\end{figure}

We took near-IR J-band spectra (1.1-1.4 $\mu$m) of OTS\,44 with 
SINFONI at the VLT at a medium spectral resolution 
(R=$\lambda$/$\Delta \lambda \sim2000$, \citealt{2013A&A...555A.107B}).
We discovered a strong, broad, and variable Paschen\,$\beta$ (Pa\,$\beta$) emission line of OTS\,44 in these spectra
\citep{2013A&A...558L...7J}, as shown in Fig.\,\ref{fig:pabeta} (left, middle). 
Furthermore, a prominent H$\alpha$ emission line is visible
in the optical spectrum of \citet{2007ApJS..173..104L} as shown in the right panel of Fig.\,\ref{fig:pabeta}. 
Both of these Hydrogen emission lines exhibit a broad profile 
with velocities of $\pm$200\,km\,s$^{-1}$ or more.
We determined the equivalent width (EW) of both lines.
The H$\alpha$ line has a symmetrically shaped profile with an EW of -141\,{\AA}, 
demonstrating that OTS\,44 is actively accreting.
The profile of the Pa\,$\beta$ line is significantly variable between the two observing epochs 
separated by a few days (EW of -7 and -4\,\r{A}). 

We determined the mass accretion rate of OTS\,44 based on the H$\alpha$ line 
by assuming
that the H$\alpha$ emission is entirely formed by accretion processes. We 
derived an H$\alpha$ line luminosity log\,$L_{H\alpha}$ ($L_{\odot}$) of 
-6.16 for OTS\,44 and applied
the empirical relation between the H$\alpha$ line and 
accretion luminosity of Fang et al. (2009), which is based on 
log\,$L_{H\alpha}$ ($L_{\odot}$) ranging between -1 and -6.
We estimate an mass accretion rate of OTS\,44 of
7.6$\times 10^{-12}$\,$M_{\odot}\,\mathrm{yr}^{-1}$.

\begin{figure}[h]
\centering
\includegraphics[width=0.6\linewidth]{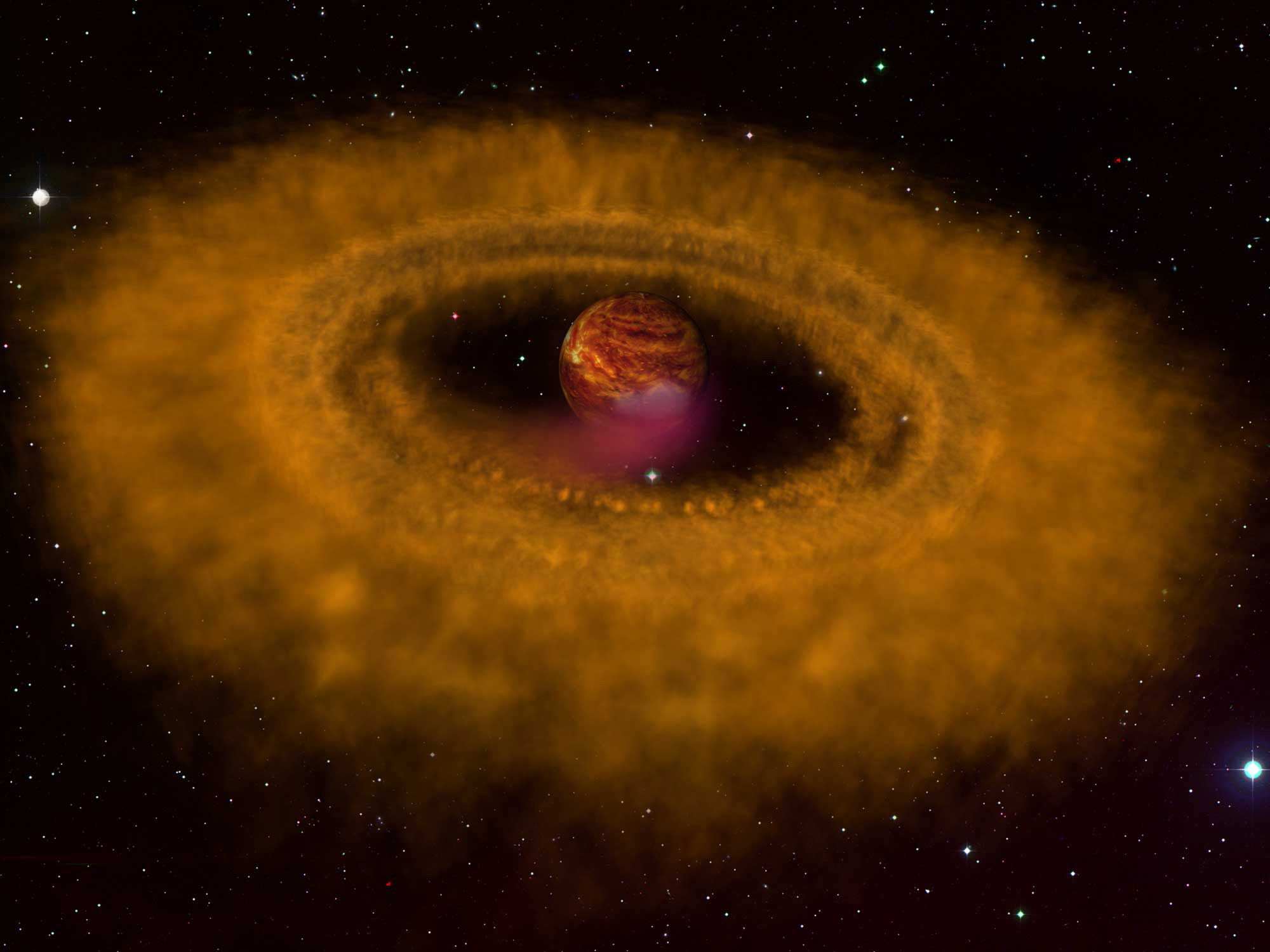}
\caption{
Artist's view of the accreting and disk-bearing free-floating planetary mass object OTS\,44. Credit: MPIA\,/\,A. M. Quetz.
}
\end{figure}

\section{Conclusions}

We have discovered strong, broad, and variable Pa\,$\beta$ emission of 
the young very low-mass substellar object OTS\,44 (M9.5)
in VLT/SINFONI spectra, which is evidence for active accretion of a planetary mass object \citep{2013A&A...558L...7J}.
We determined the properties of the disk that surrounds OTS\,44 through 
\texttt{MC3D} radiative transfer modeling of flux measurements from the optical to the far-IR including Herschel data
\citep{2013A&A...558L...7J, {joergens2014}}.
We found that OTS\,44 has a highly flared disk ($\beta>$1.3) with a mass of 
3.25$\times10^{-5}\,M_{\odot}$, i.e. about $10\,M_{Earth}$. 
We also investigated the H$\alpha$ line of OTS\,44 in a MAGELLAN\,I/IMACS
spectrum spectrum 
and 
found strong H$\alpha$ emission with an EW of -141\,{\AA} indicative of active accretion.
%
\begin{figure}[h]
\centering
\includegraphics[width=0.7\linewidth]{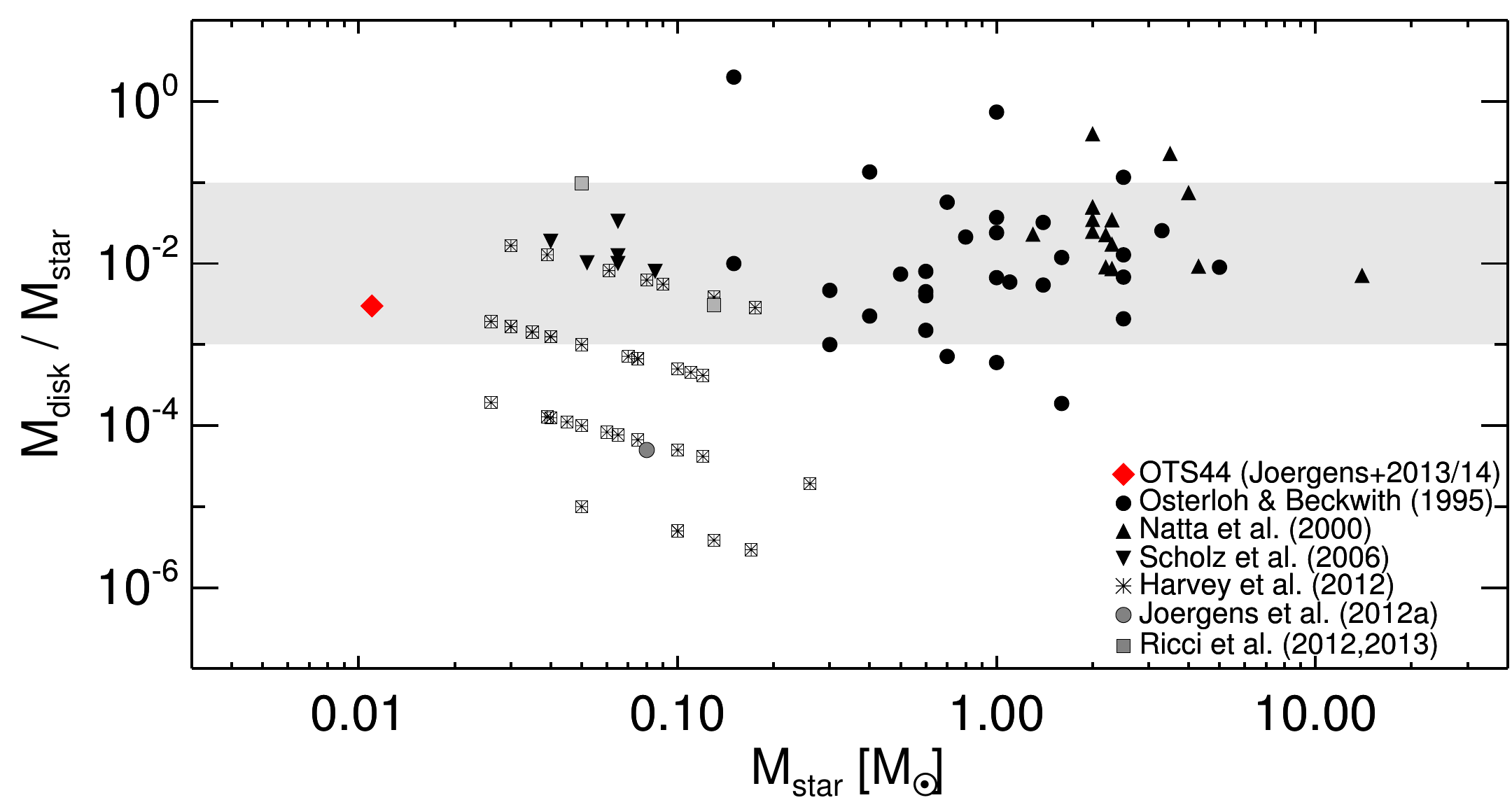}
\caption{
\label{fig:mdisk}
Relative disk mass versus central mass of stars and brown dwarfs including 
OTS\,44 (red diamond). 
Slightly revised version of similar plot from \citet{2013A&A...558L...7J}. 
}
\end{figure}
%
\begin{figure}[b]
\centering
\includegraphics[width=0.7\linewidth]{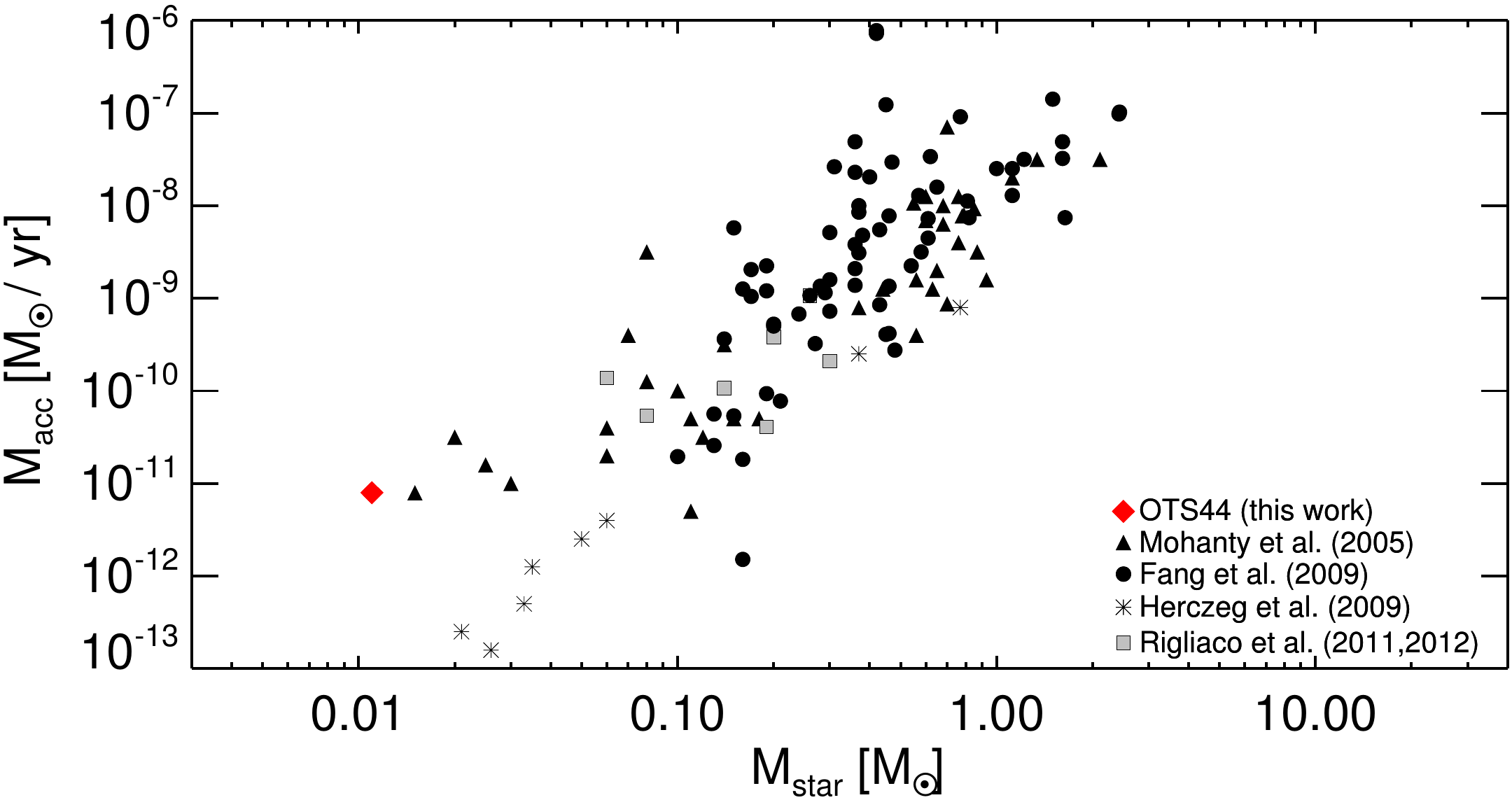}
\caption{
\label{fig:macc}
Mass accretion rate versus central mass of stars and brown dwarfs
including OTS\,44 (red diamond). From \citet{2013A&A...558L...7J}. 
}
\end{figure}
%
\begin{figure}[h]
\centering
\includegraphics[width=0.9\linewidth]{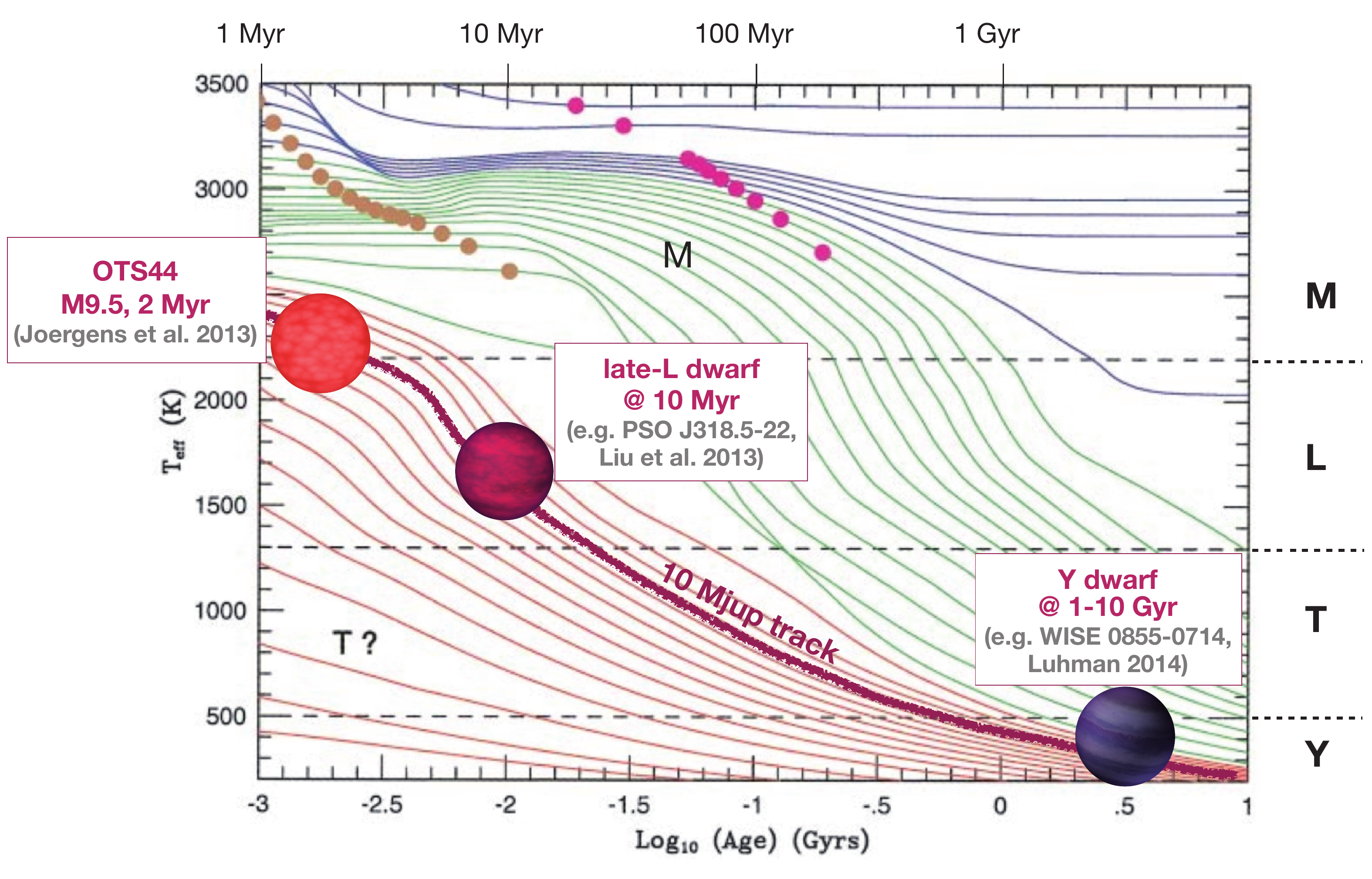}
\caption{
\label{fig:teffevol}
OTS\,44 will become a Y dwarf in $10^9$ years.
For this illustration of the evolution of the effective temperature of a very low-mass substellar object we used  
a plot of \citet{2001RvMP...73..719B}.
Note that the model temperatures are
hotter than the observationally determined temperatures, which are
1700\,K for OTS\,44, 1200\,K for PSO\,J318.5-22, and 250\,K for WISE\,0855-0714.
}
\end{figure}
%
Both the Pa\,$\beta$ and H$\alpha$ emission lines of OTS\,44 have broad
 profiles 
with the wings extending to velocities of about $\pm$200\,km\,s$^{-1}$.
The Pa\,$\beta$ emission is significantly variable on timescales of a few days, 
indicating variability in accretion-related processes of OTS\,44.
We estimated the mass accretion rate of OTS\,44 to 
7.6$\times 10^{-12}$\,$\mathrm{yr}^{-1}$
by using the H$\alpha$ line.
A mass accretion rate based on the Pa\,$\beta$ line gives a significantly higher value, and 
we speculate that part of the Pa\,$\beta$ emission might come from other processes related 
to accretion, such as outflows.
Furthermore, in the course of studying OTS\,44, we fitted a photospheric BT-Settl model to its
optical and near-IR SED and derived a lower effective temperature and higher extinction than was 
previously found \citep{2007ApJS..173..104L}.

We have presented the first detection of Pa\,$\beta$ emission for a free-floating object below the deuterium-burning limit.
Our analysis of Pa\,$\beta$ and H$\alpha$ emission of OTS\,44 demonstrates that free-floating objects of a few Jupiter masses 
can be active accretors.
OTS\,44 can be seen as free-floating analog of recently detected accreting planetary mass companions that orbit stars
(e.g., \citealt{2011ApJ...743..148B}, \citealt{2014ApJ...783L..17Z})
and it plays a key role in the study of disk evolution and accretion physics 
in an extremely low-gravity and -temperature environment.
Furthermore, OTS\,44 (M9.5) is the lowest-mass object to date for which the disk mass is determined based on far-IR data.
Also interesting in this context is the mm-detection of an M9 dwarf
\citep{2006ApJ...645.1498S}.
Our detections therefore extend the exploration of disks and accretion during the T~Tauri phase   
down to the planetary mass regime.
Plotting the relative disk masses of stars and brown dwarfs 
including OTS\,44 (Fig.\,\ref{fig:mdisk}) shows
that the ratio of the disk-to-central-mass of about 10$^{-2}$ found for 
objects between 0.03\,$M_{\odot}$ and 14\,$M_{\odot}$ is also valid for OTS\,44 at a mass of about 0.01\,$M_{\odot}$.
Furthermore,
the mass accretion rate of OTS\,44 is consistent with a decreasing
trend from stars of several solar masses to substellar objects down to 0.01\,$M_{\odot}$ (Fig.\,\ref{fig:macc}). 
It is also obvious from this figure that OTS\,44
has a relatively high mass accretion rate considering its small mass.
These observations show that the processes that accompany canonical star formation,
disks and accretion, are present down to a central mass of a few Jupiter masses. 
This suggests that OTS\,44 is formed like a star and that
the increasing number of young free-floating planets and ultra-cool  
T and Y field dwarfs are the low-mass extension of the stellar population. 

Figure\,\ref{fig:teffevol} illustrates how a very young free-floating planetary mass object of 
about 10 Jupiter masses, such as 
OTS\,44, cools down while aging and becomes an L dwarf at about 
10\,Myr, similar to PSO\,J318.5-22, and a Y dwarf at $\geq$1\,Gyr,
similar to WISE\,0855-0714. This illustration uses a plot of the evolution of the effective temperature 
by \citet{2001RvMP...73..719B}. While this might not be the most sophisticated evolutionary model to date
and discrepancies with observations can be seen in the temperature (see caption of Fig.\,\ref{fig:teffevol}),
it nevertheless shows very nicely the general picture of a substellar object shifting
through spectral classes during its evolution.

\acknowledgments{We thank K. Luhman for providing the optical spectrum of OTS\,44
and the ESO staff at Paranal for executing the SINFONI observations
in service mode.}

\end{document}